\documentclass[final,twocolumn,twoside,times,5p]{elsarticle}
\biboptions{sort&compress}
\makeatletter
\def\ps@pprintTitle{%
 \let\@oddhead\@empty
 \let\@evenhead\@empty
 \let\@oddfoot\@empty
 \let\@evenfoot\@empty
}

\usepackage[utf8]{inputenc}
\usepackage[english]{babel}
\usepackage{amssymb,amsthm,amsmath,amstext,amsbsy,amsopn}
\usepackage{bbm}
\usepackage{graphicx}
\usepackage{isotope}
\usepackage{float}
\usepackage[normalem]{ulem}
\usepackage[dvipsnames]{xcolor}
\usepackage[T1]{fontenc}
\usepackage{subcaption}
\usepackage{lipsum}
\usepackage{gensymb}
\usepackage[normalem]{ulem}

\usepackage[numbers]{natbib}  
\usepackage[colorlinks=true,citecolor=blue,linkcolor=blue,urlcolor=blue]{hyperref}
\usepackage{orcidlink}

\newcommand{\NN}{\text{NN}}
\newcommand{\NNN}{\text{3N}}

\newcommand{\ab}{{\it ab~ initio}}
\makeatletter
\def\ps@pprintTitle{%
 \let\@oddhead\@empty
 \let\@evenhead\@empty
 \let\@oddfoot\@empty
 \let\@evenfoot\@empty
}
\makeatother

\begin{document}

\begin{frontmatter}




\title{$\NN$ and $\NNN$ Tensor Force  in the $N=34$ Shell Evolution: An {\it Ab Initio} Perspective}


\author{Anil Kumar\orcidlink{0000-0003-3299-0435}\corref{cor1} }
\ead{anil@nucl.ph.tsukuba.ac.jp}
\author{Takayuki Miyagi\orcidlink{0000-0002-6529-4164}}
\ead{miyagi@nucl.ph.tsukuba.ac.jp}
\author{Noritaka Shimizu\orcidlink{0000-0003-1638-1066}}
\ead{shimizu@nucl.ph.tsukuba.ac.jp}
\address{Center for Computational Sciences, University of Tsukuba, Tennodai, Tsukuba 305-8577, Japan}

\begin{abstract}
Shell evolution plays a vital role in understanding the nuclear shell structures across the nuclear chart. In this work, we have investigated the $N = 34$ shell structure using the state-of-the-art {\it ab-initio} valence-space in-medium similarity renormalization (VS-IMSRG) approach. Notably, we employ nucleon-nucleon ($\NN$) and three-nucleon ($\NNN$) interactions derived from chiral effective field theory and make use of the spin-tensor decomposition scheme to examine the contributions of individual interaction components. We discuss the evolution of the shell structures, which have been investigated by considering the roles of various components, including central, spin-orbit, and tensor effects of $\NN$ and $\NNN$ forces, respectively. The $N=34$ shell gap gradually decreases from $^{54}$Ca as the proton occupancy in the $\pi{0f_{7/2}}$ orbital increases, and eventually disappears in the $^{62}$Ni as a consequence of the tensor-force driven shell evolution.  Our analysis reveals that this disappearance is predominantly governed by the $\NN$ tensor force, which accounts for approximately 83$\%$, while the $\NNN$ tensor force also contributes about 17$\%$.
  
\end{abstract}




\end{frontmatter}




\section{Introduction}
\label{introduction}
The shell structure provides a foundation for understanding how complex nuclear properties emerge from the underlying interactions~\cite{Reiter2018PRC, Otsuka2020PRR}. 
In recent years, it has become a major focus of both experimental and theoretical research in nuclear physics. Magic nuclei, which have a magic number of protons and/or neutrons and are characterized by a large shell gap, play an important role. 
A fundamental question is the robustness of the standard magic numbers in neutron-rich unstable nuclei.
These exotic nuclei are key systems for studying changes in shell structure as more neutrons are added, known as ``shell evolution'', and for understanding how the shell evolution is driven by nuclear interactions. 
Recently, significant progress has been made on this topic.
Many studies report that exotic nuclei exhibit the disappearance of traditional magic numbers and that the emergence of new ones in some cases as a consequence of the shell evolution~\cite{warner2004Nat, Otsuka2006PRL, Robert2009Nat}. As a few prominent examples, many theoretical and experimental efforts have shown that the traditional neutron magic $N= 8$~\cite{IWASAKI2000PLB, Pain2006PRL, IMAI2009PLB, Krieger2012PRL, Meharchand2012PRL, MORSE2018PLB},  $N=20$~\cite{MOTOBAYASHI1995PLB, TerryPRC2008, wimmer2010PRL, Doornebal2016PRC, Lica2019PRC}, and $N=28$~\cite{Bastin2007PRL, Force2010PRL, Gade2010PRL, Longfellow2021PRC} numbers disappear, and new magic numbers like $N=14$ \cite{Becheva2006PRL}, $N=16$~\cite{Ozawa2000PRL, Hoffman2008PRL, Kanungo2009PRL, Tshoo2012PRL,  Rogers2015PRC}, $N=32$ \cite{Huck1985PRC, Weinholtz2013Nat, Rosenbusch2015PRL}, or $N=34$~\cite{Steppenbeck2013Nat, Michimasa2018PRL, Liu2019PRL, Limura2023PRL} emerge in the neutron-rich regions. Nearly two decades ago, Otsuka \textit{et al.} \cite{Otsuka2001PRL, Otsuka2005PRL} identified that the shell evolutions are attributed to the monopole component of the tensor force in the employed nucleon-nucleon (NN) interactions.

Such appearance or vanishing of magic nuclei has a significant impact on various fields, including experimental and theoretical nuclear structure, as well as in astrophysics, fundamental symmetries, and interactions \cite{ARNOULD200797, Otsuka2020PRR}.
Calcium isotopes $^{40}$Ca and $^{48}$Ca have long been traditionally recognized as doubly magic nuclei. The development of radioactive ion-beam (RIB) facilities has led to the predictions of additional doubly magic nuclei of calcium isotopes, including $^{36}$Ca~\cite{Lalanne2023PRL}, $^{52}$Ca~\cite{Huck1985PRC, Weinholtz2013Nat}, and $^{54}$Ca~\cite{Steppenbeck2013Nat, Michimasa2018PRL}, which exhibit new magic numbers at N=16, N = 32, and N=34, respectively. 
Within the tensor-force driven shell evolution~\cite{Otsuka2001PRL, Otsuka2005PRL}, the $N=32$ and $N=34$ shell gaps were predicted to arise from the weakened proton-neutron attraction as protons are removed from the $\pi{0f_{7/2}}$ orbital in going from $^{62}$Ni to $^{54}$Ca. This decrease in the proton occupancy of $\pi{0f_{7/2}}$ orbital increases the energy gap between the neutron single-particle orbitals $\nu{0f_{7/2}}$ and $\nu{0f_{5/2}}$. 
The $N=34$ subshell gradually diminishes when protons are added to $^{54}$Ca and disappears in $^{60}$Fe and heavier isotones. Many theoretical attentions has been paid to investigating the $N=32$ and $N=34$ shell structures \cite{Otsuka2001PRL, Otsuka2005PRL, SMIRNOVA2010109, Sahoo2025PRCL, Yuan2025CPC}. In Ref.~\cite{Smirnova2012PRC},  a rather phenomenological way of spin-tensor decomposition on the effective $\NN$ interaction was performed to understand how the shell structure forms, but it fell short in capturing shell evolution, primarily due to the missing three-nucleon ($\NNN$) forces.

Over the past decade, the chiral effective field theory (EFT)~\cite{Epelbaum2009RMP, MACHLEIDT2011PR, Hammer2020RMP, CORAGGIO2024PPNP} has established a systematic approach to constructing the $\NN$ and $\NNN$ forces. 
These modern chiral EFT $\NN$ and $\NNN$ forces have been extensively utilized in many-body calculations to reveal the role of $\NNN$ force, including scattering process~\cite{Hupin2013PRC, Langhammer2015PRC, Kravvaris2020PRC}, spectroscopic properties~\cite{Navratil2007PRL, Maris2013PRC, Ekstom2014PRL, Holt2013PRL, Holt2014PRC, MA2019PRC, Wang2013PRL}, neutron drip line location on oxygen isotopes~\cite{Otsuka2010PRL, Hagen2012PRL, Cipollone2013PRL, MA202013PLB, ZHANG2022PLB}.
In the above-mentioned works, the $\NNN$ forces have been found to play an important role in explaining why the magic number is non-universal \cite{Otsuka2010PRL, MA202013PLB, Uesaka2016EPJP, Otsuka2020PRR, Tsunoda2017PRC, Holt2013EPJA, Holt2013JPG} and influence the spin-orbit (SO) splitting, leading to the manifestation of nuclear shell structure.
More recently, Fukui \textit{et al.} \cite{Fukui2024PLB} have investigated the role of the chiral EFT $\NNN$ force in forming the shell structure of the lighter systems in $p$ shell. In their work \cite{Fukui2024PLB}, they have seen contributions of the different rank components of the $\NNN$ force at next-to-next-leading order (${\rm N^2LO}$) level and found that the $\NNN$ forces enhance the spin-orbit (SO) splitting. 

These modern chiral EFT $\NN$ and $\NNN$ forces have become the starting point for non-perturbative approaches to derive the effective Hamiltonians, such as shell-model coupled cluster (SMCC)~\cite{Jansen2014PRL} and valence-space in-medium similarity renormalization group (VS-IMSRG)~\cite{Tsukiyama2012, Stroberg2019ARNPS} approaches.  The VS-IMSRG successfully validates the need for $\NNN$ force to reproduce the experimental behavior of the ground states, low-lying excited states, and electromagnetic properties from the lighter to the heavier systems \cite{Stroberg2019ARNPS}. 
Recently, Sahoo \textit {et al.} \cite{Sahoo2025PRCL} performed a phenomenological way of spin-tensor decomposition on the $\ab$ VS-IMSRG-derived Hamiltonian to investigate the origin of the shell gap, finding that the tensor force plays a dominant role in diminishing the $N=34$ shell gap. 
Another work \cite{Yuan2025CPC} has also investigated the $N=32$ and $N=34$ shell structures by using the VS-IMSRG approach.  However, in these cases, it is pretty challenging to understand the contribution of $\NN$ and $\NNN$ in shell evolution. To see the effect of the individual microscopic nuclear forces $\NN$ and $\NNN$ in predicting shell structure is essential.

The present study aims to investigate the shell structure in a more $\ab$ way to see the individual effects of the $\NN$ and $\NNN$ forces. Our starting approach is to construct the $\NN$ and $\NNN$ interaction from the chiral EFT \cite{Epelbaum2009RMP, MACHLEIDT2011PR, Hammer2020RMP}. Then, we have implemented the spin-tensor decomposition scheme to investigate the contributions of the different rank components of bare $\NN$ and $\NNN$ in the formation of the shell closure. After the decomposition, the VS-IMSRG method has been employed to derive the effective shell model Hamiltonian of the different rank components.  To give the benchmark of our approach, we have chosen the $N=34$ subshell closure in $fp$-shell, where $^{54}$Ca shows the $N=34$ subshell closure, which has already been confirmed from the experimental side  \cite{Steppenbeck2013Nat}. The shell closure can be understood in terms of the single-particle orbits of protons and neutrons. This leads to a discussion of shell structure from the evolution of single-particle energies (SPEs). In this work, we will discuss the role of the pronounced reduction of the tensor-interaction effects of $\NN$ and $\NNN$ forces as the number of protons in $\pi{0f_{7/2}}$ orbits decreases, and how it plays an important role in affirming the $N=34$ subshell structure. 
 
The paper is organized as follows: the methodology used in this work is presented in Sec. \ref{sec:methodology}, and the results and discussion are provided in Sec. \ref{sec:results}. Finally, the conclusion is provided in Sec. \ref{sec:summary}.

\section{Methodology} \label{sec:methodology}

In this work, we start from the intrinsic nuclear Hamiltonian of the nucleus
\begin{equation}\label{eqn:hamils}
    H = T+V_{\rm NN}+V_{\rm 3N}
\end{equation}
with the intrinsic kinetic energy term $T$ and $\NN$ ($\NNN$) potentials $V_{\rm NN}$ ($V_{\rm 3N}$).
To deeply understand the shell structure, we employ the spin-tensor decomposition~\cite{ELLIOTT1968NPA, KIRSON1973PL}.
Regarding the $\NNN$ part, we used the numerical decomposition through the rotational group algebra, different from the analytic decomposition done in recent work~\cite{Fukui2024PLB}. 
Then, the Hamiltonian~(\ref{eqn:hamils}) can be expressed as
\begin{equation}\label{eqn:std}
    H = T + \sum_{\lambda = 0}^2 V^{(\lambda)}_{\rm NN}+\sum_{\lambda =0 }^2 V_{\rm 3N}^{(\lambda)},
\end{equation}
where the rank $\lambda =0, 1,$ and 2 denote central, spin-orbit, and tensor components of $\NN$ and $\NNN$ nuclear forces, respectively.
Since the effect of the $\lambda=3$ term is negligibly small as found in Ref.~\cite{Fukui2024PLB}, we do not discuss it in this work.

In practical calculations, we construct the reference state $|\Phi\rangle$ for our system by employing the ensemble normal ordering to better capture the 3N interaction effects between the valence particles~\cite{Stroberg2017PRL}. 
The next step is to take the normal-order for all operators with respect to $|\Phi\rangle$. The Hamiltonian (\ref{eqn:hamils}) can be written in term of normal-ordered 
\begin{equation}\label{eqn:no}
    H = E^{[0]}+f^{[1]}+\Gamma^{[2]}+W^{[3]},
\end{equation}
where  $E^{[0]}$, $f^{[1]}$, $\Gamma^{[2]}$, and $W^{[3]}$ are the normal-ordered zero-, one-, two-, and three-body parts of the Hamiltonian \cite{Tsukiyama2011PRL, HERGERT2016165, Heinz2021PRC, Miyagi2022PRC} and the superscript $^{[n]}$ denotes the n-body terms. 
The residual $3N$ term $W_{ijklmn}$ is discarded by considering well-established normal-ordered two-body (NO2B) approximation \cite{Hagen2007PRC, Roth2012PRL}
\begin{equation}\label{eqn:no2b}
    H \approx H_{\rm NO2B} = E^{[0]}+f^{[1]}+\Gamma^{[2]}.
\end{equation}
In this way, the contribution of the 3N force is efficiently captured at the normal-ordered two-body levels \cite{Roth2012PRL}.
Further, the VS-IMSRG method has been used to derive the effective shell model Hamiltonian of each component in a particular valence space.

With the IMSRG, we derive an effective Hamiltonian for a limited valence space through a continuous unitary transformation controlled by the flow parameter $s$:
\begin{equation}
    H(s) = U(s)H(0)U^{\dagger}(s).
\end{equation}
Employing the exponential ansatz $U(s) = e^{\Omega(s)}$ with anti-hermitian operator $\Omega(s)$, one can find the flow equation~\cite{Morris2015PRC}
\begin{equation}
\frac{d\Omega(s)}{ds} = \sum_{n=0}^{\infty} \frac{B_{n}}{n!} [\Omega(s), \eta(s)]^{(n)},
\end{equation}
with the Bernoulli number $B_{n}$ and following definition
\begin{equation}
\begin{aligned}
[\Omega(s), \eta(s)]^{(0)} &= \eta(s), \\
[\Omega(s), \eta(s)]^{(n)} &= [\Omega(s), [\Omega(s), \eta(s)]^{(n-1)}].
\end{aligned}
\end{equation}
The anti-hermitian operator $\eta$ is known as the generator of the transformation and is designed so that the matrix elements connecting valence and other spaces are suppressed.
In this work, we used the White-type generator with the arctangent function.
Once $\Omega(s)$ is obtained, one can transform any operator as
\begin{equation}
\label{eq:trans_op}
\mathcal{O}(s) = \sum_{n=1}^{\infty} \frac{1}{n!} [\Omega(s), \mathcal{O}(s=0)]^{(n)}.
\end{equation}
It would be worth noting that we transform $T$, $V^{(\lambda)}_{\rm NN}$, and $V^{(\lambda)}_{3N}$ with Eq.~\eqref{eq:trans_op} and discuss the contribution of each component.
Throughout this work, we use the standard two-body level approximation  VS-IMSRG(2), where all the commutators are evaluated up to the two-body level and higher-body terms are dropped.
The effect of the three-body corrections is still under investigation \mbox{\cite{Heinz2021PRC, BCHE2024PRC, stroberg2024PRC, Heinz2025PRC}}. 
In Ref.~\cite{Heinz2025PRC}, it was demonstrated that the effects of the three-body operators on the first $2^{+}$ excitation energy are significant in $^{48}$Ca but are minor in $^{52}$Ca. We have estimated the effect of triple correction using VS-IMSRG($3{\rm f}_2$), introduced in Ref.~\cite{BCHE2024PRC}, in $^{54}$Ca with 1.8/2.0 (EM) interaction. We observed that the correction to the $2_1^+$  excitation energy is within a
few percent, and the impact on the current study is expected to be small.

In the present study, the effective shell model Hamiltonian is derived in the valence space of the full $fp$-shell above the $^{40}$Ca core for both protons and neutrons. 
We employed the widely used NN+3N interaction 1.8/2.0 (EM)~\cite{Hebeler2011PRC, Simonis2016PRC}. 
The NN interaction is adopted at next-to-next-to-next-to-leading order $({\rm N^3LO})$ softened by the free-space similarity renormalization group (SRG) evolution with the momentum scale $\lambda = 1.8~{\rm fm^{-1}}$, while the 3N interaction is included at next-to-next-to-leading order $({\rm {N^2LO}})$ with momentum cutoff $\Lambda=2.0~{\rm fm^{-1}}$. 
The 1.8/2.0 (EM) nicely reproduces the ground-state energies of nuclei up to $^{132}$Sn \cite{Simonis2017PRC, Stroberg2021PRL, Miyagi2022PRC}.  To further address the more physical significance, another $\Delta{{\rm N^2LO}}_{\rm GO}(394)$ interaction from ChEFT with delta isobar excitation effects is also adopted in this work, where the NN and 3N interactions are non-locally regulated with a momentum scale of 394 MeV. 
It can also reproduce the ground states and radii up to A = 132 \cite{Jiang2020PRC}. We adopted the harmonic oscillator basis at $\hbar{\omega} = 16$ MeV with truncation $e_{\rm max} = 2n+l = 12$ and additional truncation on $3N$ force as $e_1+e_2+e_3{\leq}E_{\rm 3max} = 24$, ensuring sufficient convergence \cite{Miyagi2022PRC}.

\section{Result and discussion}\label{sec:results}

\begin{figure}[!h]
	\centering 
    \hspace{-8mm}
	\includegraphics[width=\columnwidth]{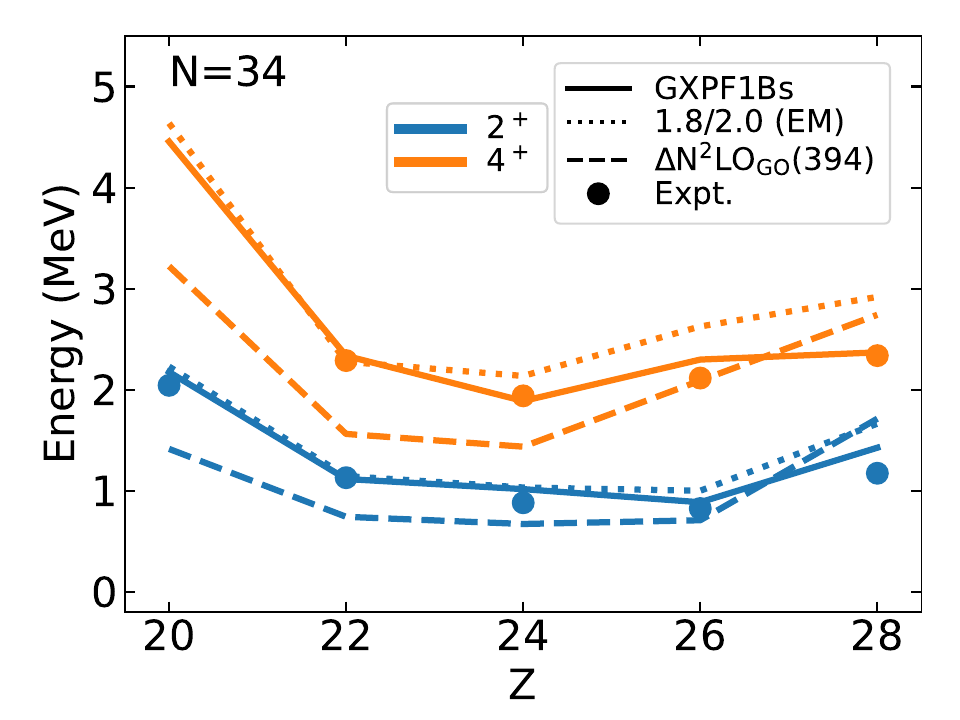}
    \caption{Excitation energies of low-lying $2^+_1$ and $4^+_1$  states of the $N = 34$ isotones  ($20\le{Z}\le{28}$) calculated within the VS-ISMRG framework using the chiral EFT interactions 1.8/2.0 (EM) and $\Delta{{\rm N^2LO}}_{\rm GO}(394)$, compared with results from the phenomenological GXPF1Bs effective interaction and experimental data \cite{nndc2024}. 
    }\label{fig:energy_2p_4p}
\end{figure}

We aim to discuss the shell evolution of $N=34$ of neutron-rich isotones with $20\le{Z}\le28$ using the modern chiral EFT-derived $\NN$ and $\NNN$ forces.
Here, we have used two types of chiral $\NN$ and $\NNN$ interactions namely 1.8/2.0 (EM) and $\Delta{{\rm N^2LO}}_{\rm GO}(394)$.
Since the first $2^{+}$ excitation energy $E_{x}(2_{1}^{+})$ is typically attributed to a large shell gap, $E_x(2_1^+)$ in even-even nuclei provides key information about the shell or subshell closures.
To ensure the predictive power of the interactions, we first calculated the $2_1^+$ and $4_1^+$ excitation energies for $N=34$ isotones with $20\leq{Z}\leq{28}$. In Fig. \ref{fig:energy_2p_4p}, we have demonstrated the computed  $E_x(2_1^+)$ and $E_x(4_1^+)$ of $N=34$ isotones with the 1.8/2.0 (EM) and $\Delta{{\rm N^2LO}}_{\rm GO}(394)$ interactions in comparison to the phenomenological GXPF1Bs \cite{Chen2019PRL} interaction and experimental data \cite{nndc2024}. The GXPF1Bs interaction reproduces the experimental $2_1^+$ and $4_1^+$ excitation energies well, as this interaction is phenomenologically fitted in this region. In the $\ab$ way, the present VS-IMSRG calculations well reproduce the systematic trend of the $E_x(2_1^+)$ energy.
Comparatively, we can see in Fig. \ref{fig:energy_2p_4p} that the 1.8/2.0 (EM) results give the better prediction of $E(2_1^+)$ than the $\Delta{{\rm N^2LO}}_{\rm GO}(394)$. The $\Delta{{\rm N^2LO}}_{\rm GO}(394)$ obtained the underestimated $E_x(2_1^+)$ and $E_x(4_1^+)$ of $^{54}$Ca and $^{56}$Ti. In the $^{54}$Ca case, the $E_x(2_1^+)$ value calculated using 1.8/2.0 (EM) differs by about 800 keV from the one obtained with $\Delta{{\rm N^2LO}}_{\rm GO}(394)$, indicating a large uncertainty arising from the underlying nuclear interactions. 
An earlier $\ab$ coupled-cluster calculation \cite{Hagen2012PRL_Ca} also predicted the first $2^+$ excitation energy in $^{54}$Ca to be 1.9 MeV, indicating a strong shell closure at N=34.  Furthermore, as shown in Fig. \ref{fig:energy_2p_4p}, the $2_1^+$ energy exhibits the $N=34$ subshell gap in $^{54}$Ca, which is diminished when increasing the proton numbers and becomes weaker in Ti, Cr, and Fe. 

From the experimental point of view, the $N=34$ subshell has already been established experimentally in $^{54}$Ca \cite{Steppenbeck2013Nat, Michimasa2018PRL} and $^{52}$Ar \cite{Liu2019PRL}.  Nevertheless, we can see that the trends of $2_1^+$ and $4_1^+$ states of $N=34$ isotones calculated using both interactions are consistent. Recently, similar calculations were performed with the VS-IMSRG method using 1.8/2.0 (EM) or ${\rm NN-N^3LO+3N_{\rm lnl}}$ interactions to examine the behavior of the $N=32$ and $N=34$ subshell gaps across the first $2^+$ energies \cite{Sahoo2025PRCL, Yuan2025CPC}.

\begin{figure*}[!h]
	\centering 
    \hspace{-8mm}
	\includegraphics[width=\columnwidth]{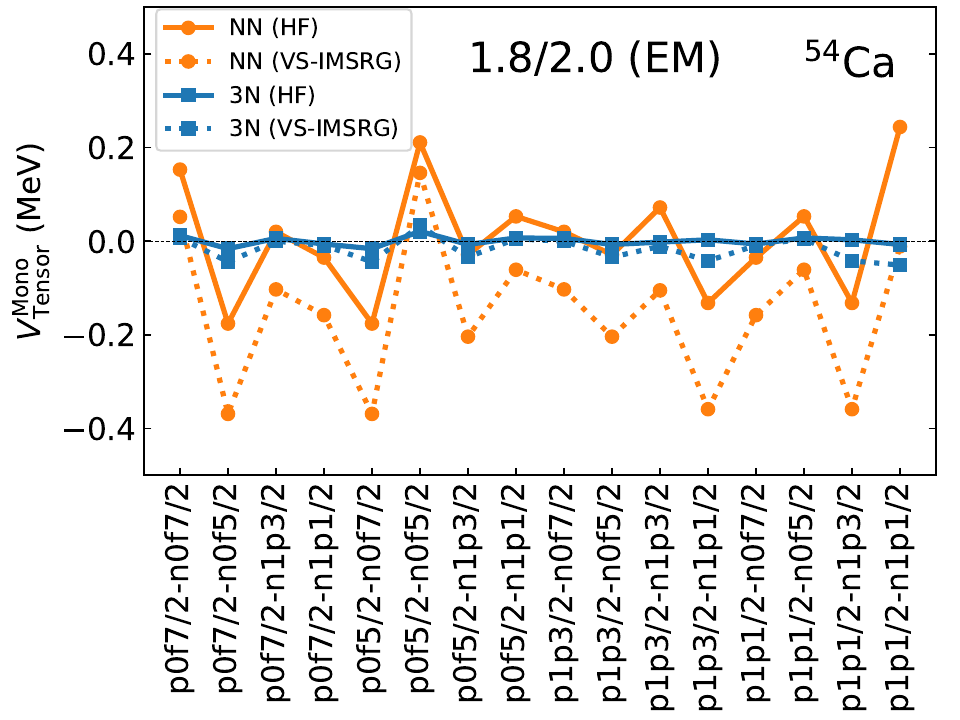}
    \includegraphics[width=\columnwidth]{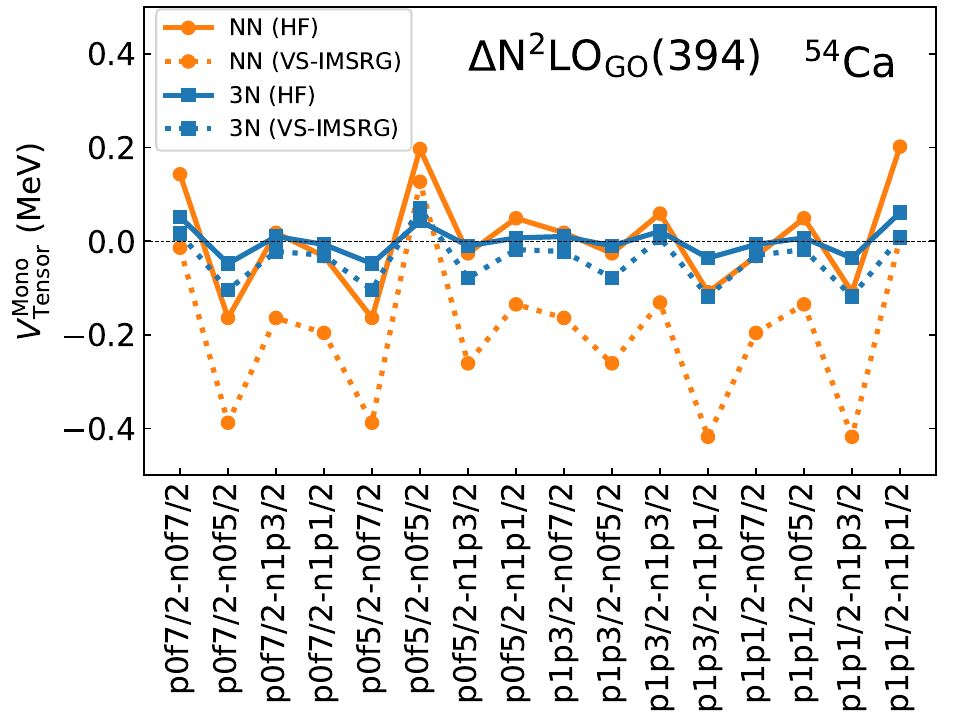}
    \caption{ Monopole strength of the tensor component of the $\NN$ and $\NNN$  interactions in $^{54}$Ca, calculated using the 1.8/2.0 (EM) interaction (left) and $\Delta{{\rm N^2LO}}_{\rm GO}(394)$ interaction (right). The dotted line labeled ``HF'' represents the results obtained from the initial Hamiltonian in the Hartree-Fock basis before IMSRG evolution, whereas solid lines labeled ``VS-IMSRG'' represent the results obtained from the valence space effective Hamiltonian after the IMSRG evolution.  } \label{fig:std_n34_ten}
\end{figure*}

The present work aims to examine how the $\NN$ and $\NNN$ forces are crucial in shaping the $N=34$ subshell structure.
To understand this, we have implemented spin-tensor decompositions of the chiral EFT-derived $\NN$ and $\NNN$ forces to examine the contributions of different-rank components, thereby providing more insight into the shell structure. As in the previous study \cite{Sahoo2025PRCL}, they have implemented the spin-tensor decomposition on the full Hamiltonian derived from the VS-IMSRG approach. However, this approach does not allow for a separate and systematic analysis of the individual contributions of the $\NN$ and $\NNN$ forces to the development of shell structure. In the present case, we first implement the spin-tensor decomposition on the bare Chiral $\NN$ and $\NNN$ interactions in Eq. (\ref{eqn:std}) and then apply the VS-IMSRG approach to obtain the effective Hamiltonian for each component, which enables a more detailed analysis of individual contributions from $\NN$ and $\NNN$.

Furthermore, we have investigated the shell evolution by considering the role of the effects from different components of realistic nuclear forces in terms of effective single-particle energies (ESPEs), which represent the nature of single-particle states and are generated by the mean field incorporating the effects from other nucleons outside the inert core. Note that the ESPEs are not observable quantities but provide insightful information about the shell structure.  Duguet et al. \cite{Duguet2012PRC} also emphasized that the ESPEs are not observable quantities because they depend on the resolution scale and representation of the nuclear Hamiltonian. In particular, they may change under the unitary transformation of the Hamiltonian, whereas true observables remain invariant. The ESPEs are calculated using the monopole component of the Hamiltonian, where the monopole components can be expressed in terms of the occupation number $\hat{n}_j$ of each $j$ orbit. The ESPEs can be expressed in terms of the single-particle state $j$, denoted by $\mathcal{E}_j$
 \begin{eqnarray}\label{eqn:espe}
   \mathcal{E}_j = \mathcal{E}_j^0+\sum_{j^\prime}{V_{jj^\prime}^{\rm (Mono)}{{\langle{N}}_{j^{\prime}}\rangle}},
\end{eqnarray}
where the $\mathcal{E}_j^0$ and  $V_{jj^\prime}^{\rm (Mono)}$ stand for the single-particle energy of $j$ orbit relative to the core and the strength of the monopole interaction between two orbits $j$ and $j^\prime$, respectively. $\langle{N_{j^\prime}}\rangle$ represents the occupation number of the $j^{\prime}$ orbit. In this study, the occupation numbers $\langle{N_{j^\prime}}\rangle$ are taken from the resulting shell-model wave functions.
The change to the ESPEs in orbits arises from the effects of monopole components of the $\NN$ and $\NNN$ interactions, which play an important role in describing the observed spectroscopic properties.   In this work, we will discuss how the ESPEs of orbit $j$ vary due to occupying the $j'$ orbit, for which we have systematically investigated the shell closures at $N=34$ for the Ca and Ni isotopes.

As discussed in Ref. \cite{NTsunoda2011PRC}, the monopole components of the tensor force are weakly or barely affected by the renormalization procedure, namely, a many-body perturbation theory type approach. In contrast to expectations, we found the strong deviation in the monopole component of the tensor for $\NN$ forces for 1.8/2.0 (EM) and $\Delta{{\rm N^2LO}}_{\rm GO}(394)$ interaction by the IMSRG evolution. The behavior of the monopole strength of the $\NN$ and $\NNN$ tensor force in $^{54}$Ca is illustrated in Fig. \ref{fig:std_n34_ten} for both interactions used in this work.  Fig. \ref{fig:std_n34_ten}, the label ``HF'' denotes the tensor-monopole components calculated from the initial Hamiltonian ($H(s=0)$) represented in the Hartree-Fock single-particle basis and normal-ordered with respect to the Hartree-Fock reference state, before the IMSRG evolution. The label ``VS-IMSRG'' denotes the corresponding tensor-monopole components calculated using the final valence-space effective Hamiltonian obtained after the IMSRG evolution and valence-space decoupling.  We can see in Fig. \ref{fig:std_n34_ten} that the monopole strength of the $\NN$ tensor force is strongly influenced by the IMSRG evolution. Moreover,  the tensor monopole component of the $\NNN$ force is also significantly affected by the IMSRG evolution. This behavior indicates that induced many-body contributions and operator renormalization in the IMSRG framework play an important role in reshaping tensor components, particularly at the monopole level. Moreover, a consistent pattern is observed in the case of $^{62}$Ni for both interactions considered. This is consistent across different nuclear systems, suggesting that the modification of the tensor monopole component is not an isolated feature, but rather a systematic effect associated with the IMSRG evolutions. A more detailed discussion on the renormalization persistency is given in \ref{sec:Mono_tensor}.

\begin{figure*}[!h]
  \centering
  \begin{subfigure}{0.49\textwidth}
  \phantomcaption\label{fig:n34_isotone_a}
    \includegraphics[width=\linewidth]{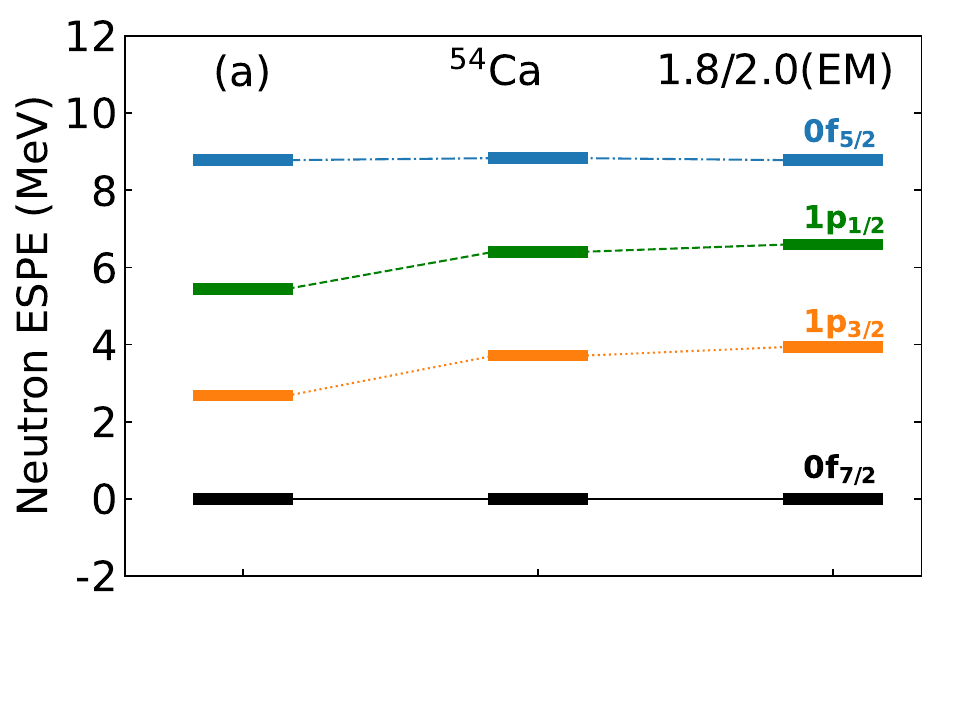}
  \end{subfigure}
  \begin{subfigure}{0.49\textwidth}
  \phantomcaption\label{fig:n34_isotone_b}
    \includegraphics[width=\linewidth]{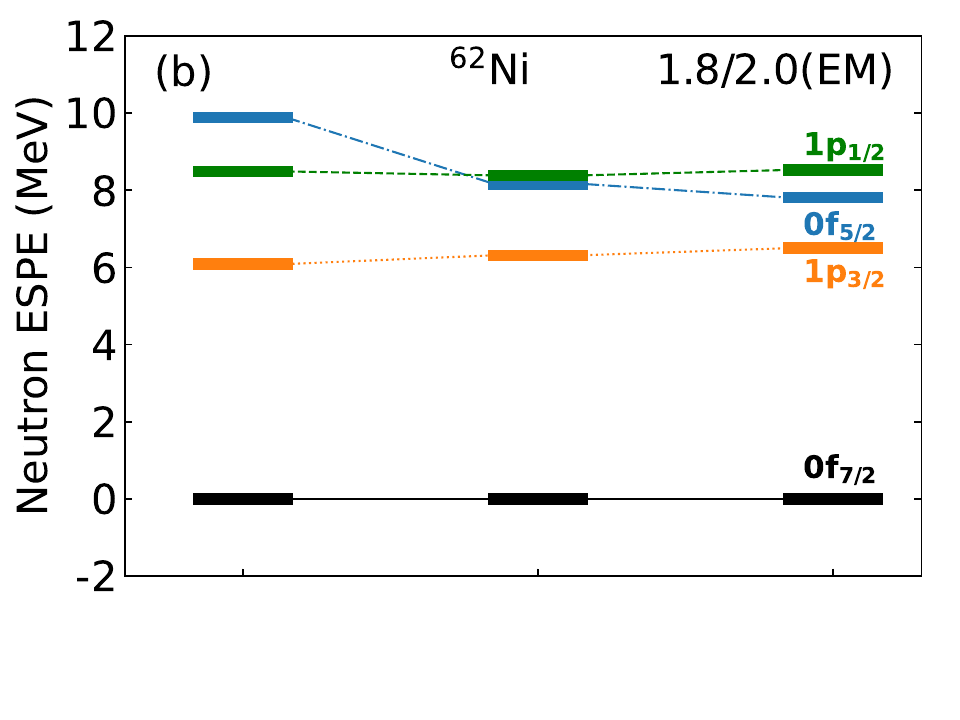}
  \end{subfigure}
  \par\medskip 
\vspace{-1.4cm}
  \begin{subfigure}{0.49\textwidth}
  \phantomcaption\label{fig:n34_isotone_c}
    \includegraphics[width=\linewidth]{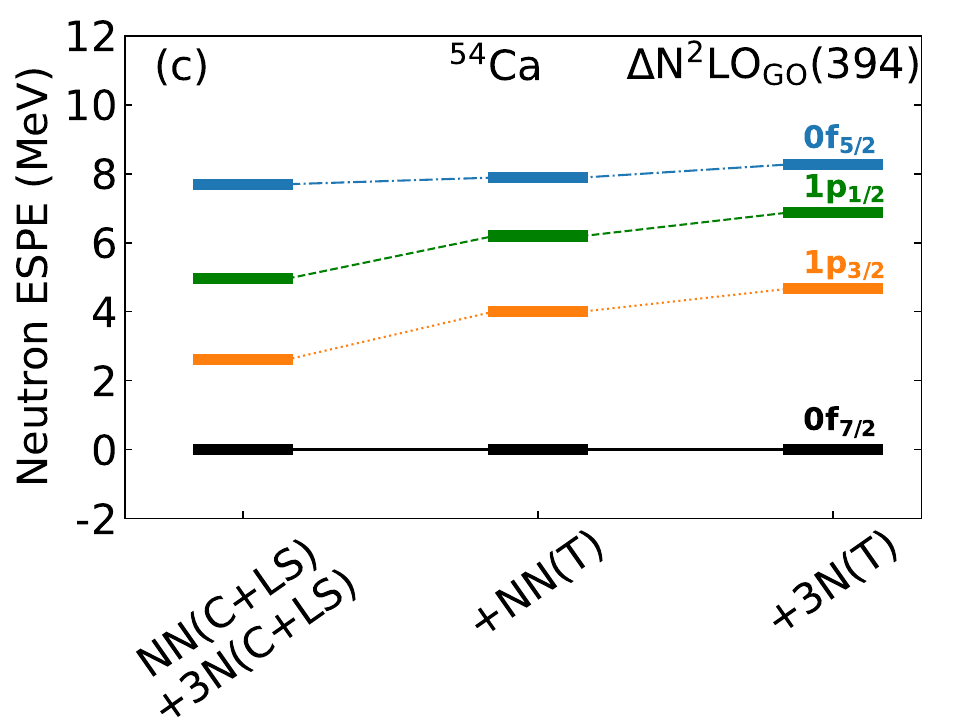}
  \end{subfigure}
  \begin{subfigure}{0.49\textwidth}
  \phantomcaption\label{fig:n34_isotone_d}
    \includegraphics[width=\linewidth]{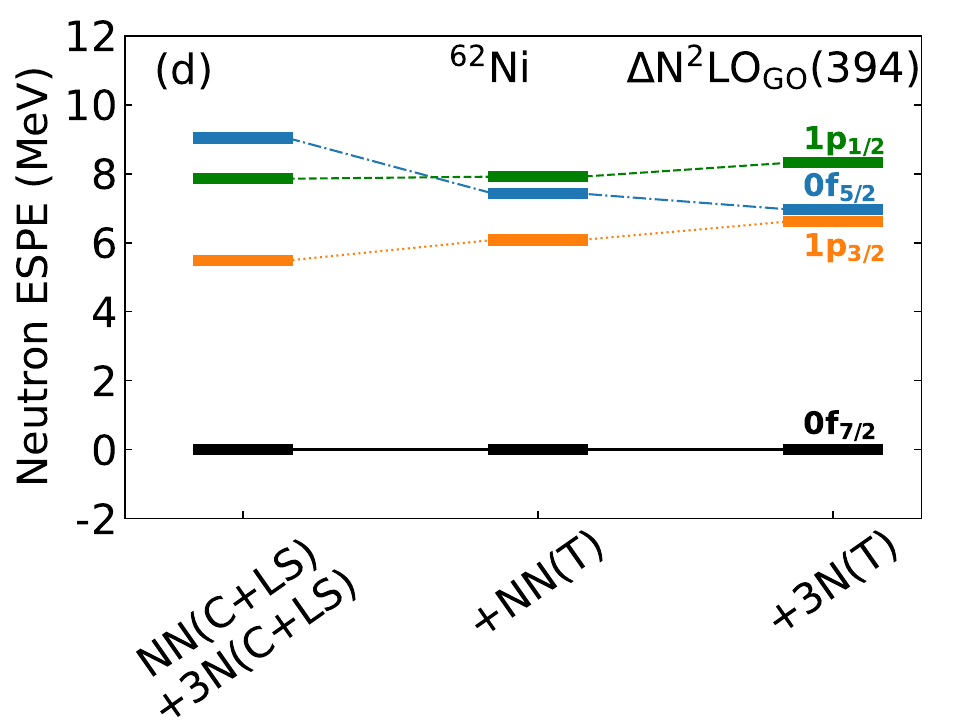}
  \end{subfigure}
  \caption{Neutron ESPEs of $^{54}$Ca ($Z=20$) and $^{62}$Ni ($Z=28$) as a function of spin-tensor decomposition components of $\NN$ and $\NNN$ force. Results are shown for two chiral interactions: (a) $^{54}$Ca and (b) $^{62}$Ni obtained with the 1.8/2.0 (EM) interaction; (c) $^{54}$Ca and (d) $^{62}$Ni obtained with the  $\Delta{{\rm N^2LO}}_{\rm GO}(394)$ interaction. The individual contributions from the $\NN$ and $\NNN$ tensor components are displayed separately to illustrate their relative roles in shaping the $N=34$ shell structure.}
  \label{fig:n34_isotone}
\end{figure*}

 In Figure \ref{fig:n34_isotone}, we have demonstrated the calculated ESPEs as a function of spin-tensor decomposition components of $\NN$ and $\NNN$ force for $^{54}$Ca ($Z=20$) and $^{62}$Ni ($Z=28$). 
As illustrated in Figs.~\ref{fig:n34_isotone_a} and \ref{fig:n34_isotone_c}, for $^{54}$Ca, the neutron $\nu{0f_{5/2}}$ orbital appears above the neutron $\nu{1p_{1/2}}$ orbital, leading to the emergence of the $N=34$ subshell gap between them. 
We note that this is observed in both 1.8/2.0 (EM) and $\Delta{{\rm N^2LO}}_{\rm GO}(394)$ results.
Shell or subshell closures are initially characterized by the excitation energy of the first $2^+$ state. Experimentally, the emergence of the $N=34$ subshell closure was established based on the first $2^+$ excitation energy by Steppenbeck et al. \cite{Steppenbeck2013Nat}. They observed the first $2^+$ excitation energy to be 2.043 MeV, a relatively large value that provides strong evidence for the subshell closure at $N=34$. The present calculations of the first $2^+$ energy using the 1.8/2.0 (EM) interaction also reproduce this value reasonably well, which can be seen in Fig. \ref{fig:energy_2p_4p}.  The calculations for $^{54}$Ca with both interactions indicate that the $2_1^+$ state primarily arises from neutron particle-hole excitation across the $N=34$ subshell gap. Notably, our results show that, although the $E_x(2^+)$ for $^{54}$Ca is comparable to the $\nu{0f_{5/2}}-\nu{1p_{1/2}}$ single-particle energy gap at $N=34$, the underlying structure of the state is governed by the correlated neutron excitations.

\begin{figure*}
	\centering 
    \hspace{-8mm}
	\includegraphics[width=\columnwidth]{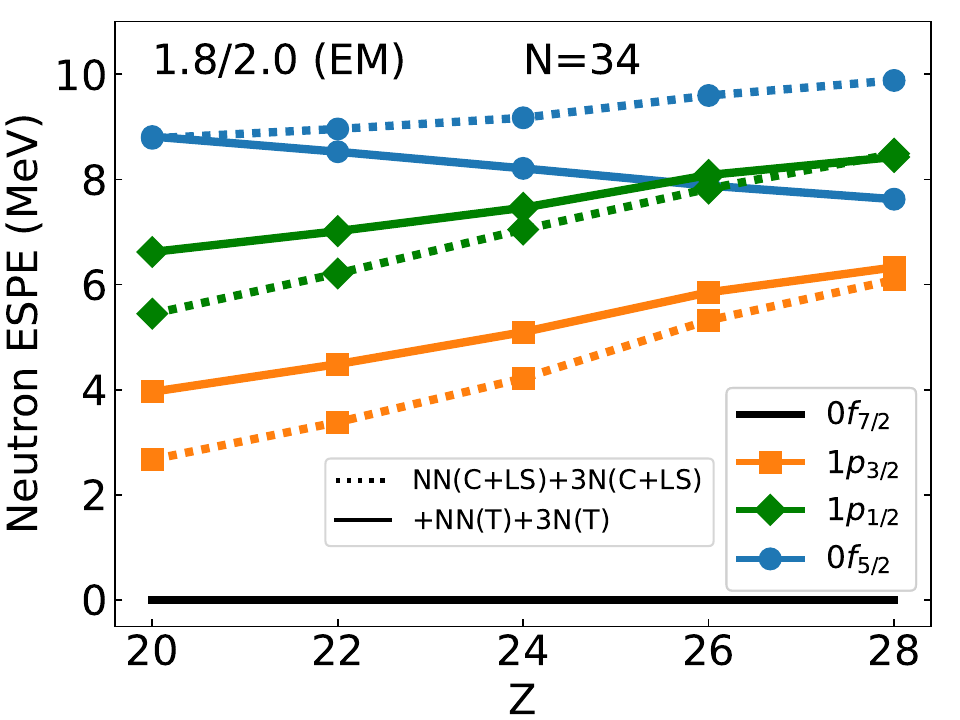}
    \includegraphics[width=\columnwidth]{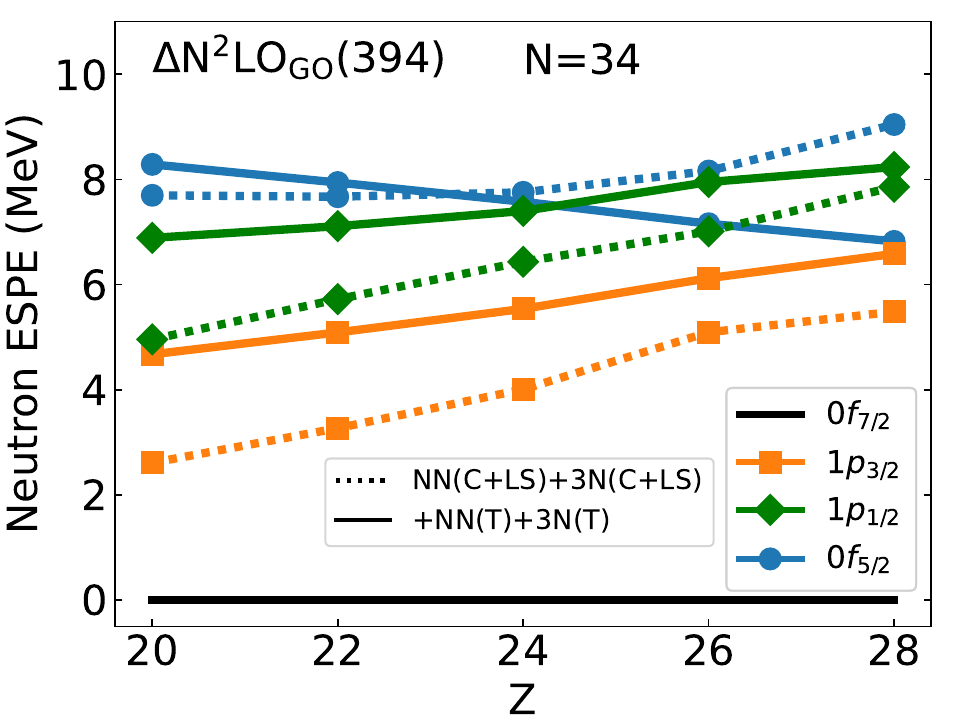}
    \caption{Systematically calculated ESPEs for $N=34$ isotonic chain from $Z=20$ to $Z=28$, shown for the 1.8/2.0 (EM) (left panel) and $\Delta{{\rm N^2LO}}_{\rm GO}(394)$ (right panel) interactions. Dotted lines denote calculations excluding the tensor components of the $\NN$ and $\NNN$ force, whereas solid lines show including their tensor contributions.} \label{fig:std_n34}
\end{figure*}
 
The $N=34$ subshell closure in $^{54}$Ca arises from the absence of the tensor-force attraction between $\pi{0f_{7/2}}$ and $\nu{0f_{5/2}}$ orbitals, as the $\pi{0f_{7/2}}$ proton orbital remain unoccupied. As the occupancy of protons in the $0\pi{f}_{7/2}$ orbital increases, the single particle level structure of $^{62}$Ni evolves, as the energy gap between spin-orbit partners orbitals $\nu0f_{7/2} ~{\rm and}~\nu{0f_{5/2}}$ is reduced due to the tensor-force effect, which can see in Figs.~\ref{fig:n34_isotone_b} and \ref{fig:n34_isotone_d}. 
Again, the physics picture does not depend on the choice of the interactions examined here.
This reduction of the spin-orbit partner in $^{62}$Ni can be understood within the framework of the mechanism presented in Ref.~\cite{Otsuka2020PRR}.  Since the ${0f_{7/2}}$ and ${0f_{5/2}}$ orbitals correspond to $j_> = (l+1/2)$ and $j_{<}=(l-1/2)$ types, respectively. Therefore, the proton-neutron monopole interaction between  $\pi{0f_{7/2}}-\nu{0f_{7/2}}$ ($\pi{0f_{7/2}}-\nu{0f_{5/2}}$) arising from the tensor force is repulsive (attractive) in nature.
The computed monopole strengths for tensor components of $\NN$ and $\NNN$ forces for $^{54}$Ca have been shown in Fig. \ref{fig:std_n34_ten} as a function of proton-neutron orbital configurations with or without an IMSRG evolution procedure. Figure \ref{fig:std_n34_ten} demonstrates that the IMSRG evolution strongly affected the monopole strength of the tensor force. For the case of interaction between $\pi0f_{7/2}-\nu0f_{5/2}$, the monopole strength of tensor components shows a strong attractive strength for the $\NN$ and $\NNN$ interactions. However, the $\NN$ monopole tensor component exhibits a more attractive nature than $\NNN$ monopole tensor component. As a result, occupying protons in the $\pi0f_{7/2}$ orbital, the $\NN$ tensor force strongly reduces the gap between neutron spin-orbit partner $\nu{0f_{7/2}}$ and $\nu{0f_{5/2}}$ orbitals in $^{62}$Ni and shows dominance in bringing down $\nu{0f{5/2}}$ orbit. Moreover, the $\NNN$ tensor force also plays a significant role in lowering the $\nu{0f_{5/2}}$ in $^{62}$Ni. Overall, the present investigation indicates that the disappearance of the $N=34$ shell gap is predominantly governed by the $\NN$ tensor force, which contributes about 83$\%$, while the $\NNN$ tensor force also contributes about 17$\%$.

To gain deeper insights into the role of the $\NNN$ tensor force contributions in the $N=34$ shell evolutions, we further analyze the individual components of the initial $\NNN$ interactions in the initial Hamiltonian. The starting Hamiltonian contains the chiral $\NN$ force at  ${\rm N^3LO}$ and chiral $\NNN$ force at ${\rm N^2LO}$ level. The chiral $\NNN$ force includes the three contributions: the two-pion ($2\pi$) exchange term, one-pion ($1\pi$) exchange term, and the contact term (ct) \cite{MACHLEIDT2011PR, Fukui2024PLB}.  
The ct terms of $\NNN$ force do not depend on the spin and momenta, and thus the ct term can affect only the central force, not the tensor force. Here, we have numerically confirmed that the ct term does not contribute to $\NNN$ tensor force. 
We found that the $2\pi$-exchange $\NNN$ tensor force significantly changes the shell structure. Furthermore, the $2\pi$ exchange term can be decomposed into components proportional to the low-energy constants $c_1$, $c_3$, and $c_4$~\cite{MACHLEIDT2011PR, Machleidt_2016}. 
Further investigation of the $2\pi$- exchange term clarified that the $c_3$ LEC term plays a major role in the $\NNN$ tensor force.

Finally in Fig. \ref{fig:std_n34}, we present the systematic behavior of neutron ESPEs along the $N=34$ isotonic chain with $20\le{Z}\le{28}$, obtained using the 1.8/2.0 (EM) and $\Delta{{\rm N^2LO}}_{\rm GO}(394)$ interactions. These figures illustrate the evolution of neutron ESPEs for single-particle levels of $fp$-shell nuclei, ranging from $^{54}$Ca to $^{62}$Ni. 
As shown in the Fig. \ref{fig:std_n34}, the $\NN$ and $\NNN$ tensor forces lowers the $\nu{0f_{5/2}}$ orbital, placing it between $\nu{1p_{1/2}}$ and $\nu{1p_{3/2}}$ orbitals in $^{60}$Fe, and $^{62}$Ni as a proton occupancy of  $\pi{0f_{7/2}}$ orbital increases. Consequently, the $N=34$ subshell gap is reduced and eventually disappears. In contrast, for $^{54}$Ca, where the tensor-force effect is weaker, which shifts the $\nu{0f_{5/2}}$ above the $\nu{1p_{1/2}}-\nu{1p_{3/2}}$ energy gap, leading to the emergence of $N=34$ subshell closure. The results obtained so far are consistent with previous studies based on phenomenological interactions, such as GXPF1Br \cite{Steppenbeck2013Nat}, which reasonably describes shell evolution in this region \cite{Otsuka2020PRR}.

\section{Summary and conclusions}\label{sec:summary}
We investigate shell evolution driven by the tensor component of the nuclear force derived from the chiral effective field theory. In this work, we perform a detailed analysis of shell evolution of neutron-rich $N=34$ isotopes from Ca to Ni using the VS-IMSRG framework, employing chiral $\NN$ and $\NNN$ interaction, namely the 1.8/2.0 (EM) and $\Delta{{\rm N^2LO}}_{\rm GO}(394)$. The calculated first ${2}^+$ energies of $N=34$ isotones are in good agreement with available experimental data and phenomenological GXPF1Bs interaction. We also find that the $E(2^+_1)$ energy in  $^{54}$Ca is comparable to the $\nu{0f_{5/2}}- \nu{1p_{1/2}}$ gap at $N=34$, indicating that the underlying structure of the $2^+_1$ state is governed by the correlated neutron excitations.  

To discuss the $N=34$ shell structure, we introduce the spin-tensor decomposition of the $\NN$ and $\NNN$ interactions, enabling a systematic investigation of the shell evolution across the neutron-rich $N=34$ isotonic chain. Starting from the decomposed components of the nuclear force, we derive the effective shell-model Hamiltonian in the $fp$ valence space using the VS-IMSRG framework. This approach allows us to isolate and quantify the contribution of tensor components to the evolution of shell structure.

Our analysis demonstrates that tensor components of both $\NN$ and $\NNN$ forces play an important role in shaping the shell evolution. In particular, we find that these tensor forces play a prominent role in disappearing the $N=34$ subshell closure in $^{62}$Ni. The $\NN$ tensor force strongly modifies the spin-orbit splitting between the $\nu{0f_{7/2}}$ and $\nu{0f_{5/2}}$ orbitals, leading to the erosion of both the $N=32$ and $N=34$ subshell closure. Moreover, the $\NNN$ tensor force provides a further reduction of approximately $\sim 17\%$ in the $\nu{0f_{7/2}}- \nu{0f_{5/2}}$ gap. We also point out that a large fraction of the $\NNN$ tensor force comes from the $2\pi$ exchange term. We note that the above findings do not depend on the $\NN$+$\NNN$ interaction choices tested here.

As discussed in  \cite{ytsunoda2013PRC, Anil2026PLB}, changes in single particle energies play a crucial role in determining all essential features of nuclear structure, such as deformation. 
In conclusion, the present study provides new insights into the evolution of the shell structure in terms of the effective single particle energies in neutron-rich $N=34$ isotones, highlighting the crucial role of tensor components of chiral $\NN$ and $\NNN$ interactions within an {\it ab-initio} framework. These results underscore the importance of a microscopic understanding of nuclear forces describing the shell evolution.

\section*{Acknowledgments}
We would like to thank Tokuro Fukui for the fruitful discussion.
We acknowledge the support of the ``Program for promoting research on the supercomputer Fugaku'', MEXT, Japan (JPMXP1020230411), JST ERATO Grant No. JPMJER2304, Japan, and KAKENHI (25K00995, 25H01268, 25K07294 and 25K07330). We used the Nuhamil code \cite{miyagi2023epja} and IMSRG++ code \cite{Stroberg2017PRL, imsrg++} to prepare the effective Hamiltonian.
The shell model Hamiltonian diagonalizations are performed using the KSHELL code \cite{SHIMIZU2019372} mainly on the Fugaku supercomputer at R-CCS, RIKEN (hp240213,hp250224), and the Pegasus supercomputer at the University of Tsukuba (MCRP program: NUCLSM and wo22i002).

\section*{Data Availability}
The data are available from the authors upon reasonable request.

\appendix
\renewcommand{\thefigure}{A\arabic{figure}}
\setcounter{figure}{0}
\begin{figure*}[!ht]
	\centering 
    \hspace{-8mm}
	\includegraphics[width=\columnwidth]{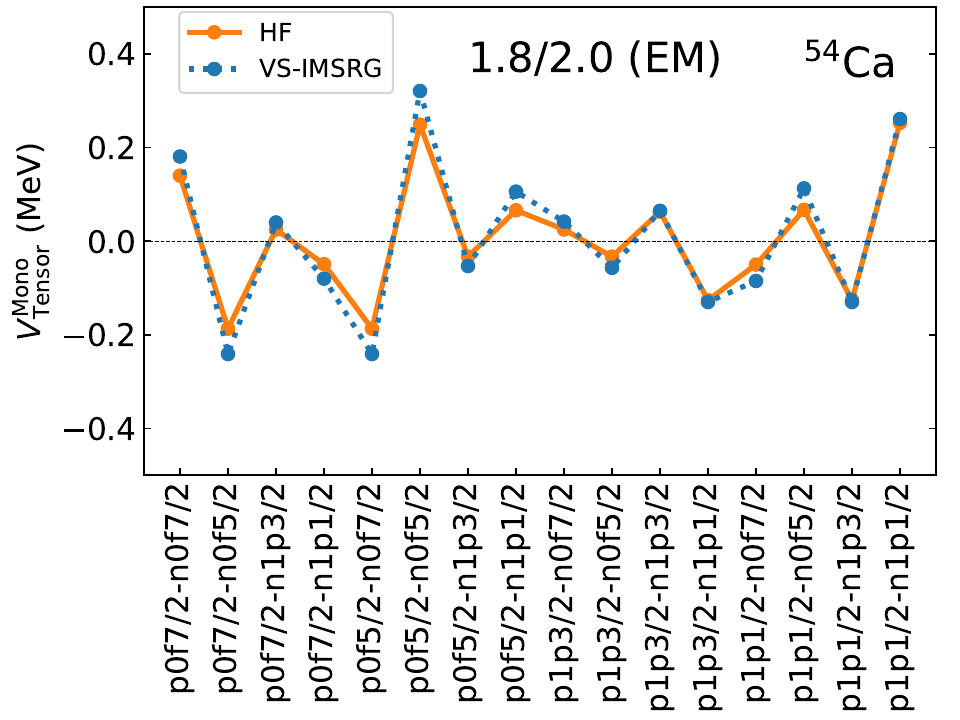}
    \includegraphics[width=\columnwidth]{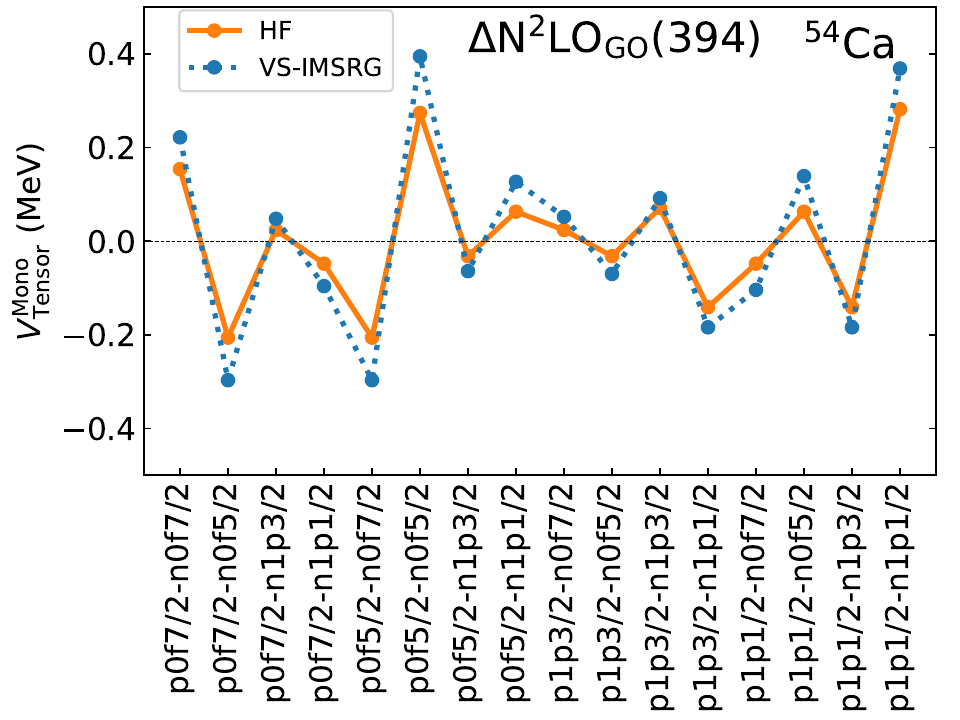}
    \caption{ Monopole strength of the tensor component of the NN and 3N  interactions in $^{54}$Ca, calculated using the 1.8/2.0 (EM) interaction (left) and $\Delta{{\rm N^2LO}}_{\rm GO}(394)$ interaction (right). The dotted line labeled ``HF'' represents the results obtained from the initial Hamiltonian in the Hartree-Fock basis before IMSRG evolution, whereas solid lines labeled ``VS-IMSRG'' represent the results obtained from the valence space effective Hamiltonian after the IMSRG evolution. The tensor monopole strengths are computed using the spin-tensor decomposition in a phenomenological manner on the full Hamiltonian derived with and without IMSRG evolution.} \label{fig:std_n34_ten_full}
\end{figure*}

\section{Renormalization persistency of Tensor force }\label{sec:Mono_tensor}
To investigate the renormalization persistency of tensor force, we implemented the spin-tensor phenomenological decomposition on the total effective Hamiltonian obtained with and without the VS-IMSRG evolution, following a phenomenological approach similar to that adopted in the previous work \cite{Sahoo2025PRCL}. The results show that the monopole tensor strength remains nearly unchanged throughout the IMSRG evolution, indicating that the monopole component of the tensor force is preserved in the total Hamiltonian, consistent with the discussion \cite{NTsunoda2011PRC}. The obtained results with this approach are shown in the Fig. \ref{fig:std_n34_ten_full}.

As discussed in the main part of the manuscript, a different behavior is observed when the spin-tensor decomposition is applied directly to the chiral $\NN$ and $\NNN$ interactions prior to deriving the effective Hamiltonian. In this case, after performing the IMSRG evolution, the tensor monopole strength changes significantly for both the $\NN$ and $\NNN$ components. This demonstrates that the IMSRG evolution strongly renormalizes the individual tensor contributions of the bare nuclear forces.

\bibliographystyle{apsrev4-1}
\bibliography{main}

\end{document}